\documentclass{PoS}
\usepackage{amssymb,amsmath,graphicx}
\usepackage{slashed}

\title{Precision calculations of nucleon charges $g_A$, $g_S$, $g_T$}

\ShortTitle{Precision calculations of nucleon charges $g_A$, $g_S$, $g_T$}

\author{\speaker{Rajan Gupta} \footnote{LA-UR-14-29293} \\ 
	Theoretical Division, Los Alamos National Laboratory, Los Alamos, NM 87545, USA \\
	E-mail:\email{rajan@lanl.gov}}

\author{{Tanmoy Bhattacharya}\\
	Theoretical Division, Los Alamos National Laboratory, Los Alamos, NM 87545, USA \\
	E-mail:\email{tanmoy@lanl.gov}}

\author{{Anosh Joseph}\\
	John von Neumann Institute for Computing, DESY, 15738 Zeuthen, Germany \\
	E-mail:\email{anosh.joseph@desy.de}}

\author{{Huey-Wen Lin} \\
	Department of Physics, University of Washington, Seattle, WA 98195\\
	E-mail:\email{hwlin@phys.washington.edu}}

\author{{Boram Yoon} \\
	Theoretical Division, Los Alamos National Laboratory, Los Alamos, NM 87545, USA\\
	E-mail:\email{boram@lanl.gov}}

\abstract{We present a detailed analysis of statistical and systematic
  errors in the calculation of matrix elements of iso-vector scalar,
  axial and tensor charges between a neutron and a proton state. These
  analyses are being done on dynamical $N_f=2+1+1$ HISQ configurations
  generated by the MILC Collaboration using valence clover fermions.
  Using ensembles at three values of the lattice spacing
  ($a=0.12,\ 0.09,$ and $0.06$ fm) and three values of the quark mass
  ($M_\pi \approx 310,\ 220$ and $130$ MeV) we find that the estimates
  of the tensor charge are stable and it can be extracted with $5\%$
  precision with O(10,000) measurements. We also find that higher
  statistics are needed to resolve the various uncertainties in the
  calculation of $g_A$ and improve the signal in $g_S$, which with
  present data has large errors. A brief status report on the mixing
  and renormalization of novel operators contributing to nEDM is also
  given.}

\FullConference{The 32nd International Symposium on Lattice Field Theory\\
   23-28 June, 2014\\
    Columbia University New York, NY}

\begin{document}

\section{Introduction}

Precise calculations of the matrix elements of iso-scalar and
iso-vector bilinear quark operators within nucleon states are needed
to probe many exciting areas of the Standard Model (SM) and its extensions.
In Ref.~\cite{Bhattacharya:2011qm}, we showed that new scalar and
tensor interactions at the TeV scale could give rise to corrections at
the $10^{-3}$ level in precision measurements of the helicity flip
parts of the decay distribution of (ultra)cold neutrons (UCN). This
sensitivity is reachable in experiments currently under construction
and being planned. Even if these experiments see a signal, to
constrain the allowed parameter space of beyond the SM (BSM) models,
however, requires that matrix elements of isovector scalar and tensor
bilinear quark operators are known to 10--20\% accuracy. Similarly, in
Ref.~\cite{Bhattacharya:2012nEDM}, we showed that to probe novel sources of CP violation in
neutron electric dipole moment (nEDM), a combination of matrix elements
of the iso-scalar and iso-vector tensor operators are needed to estimate
the contribution of the quark EDM to the nEDM. Lattice calculations
are well poised to provide these estimates with the desired
precision. 

In these proceedings, we present a detailed analysis of statistical
and systematic errors in such calculations using 9 ensembles of 2+1+1
flavor HISQ lattices generated by the MILC
collaboration~\cite{Bazavov:2012xda}. The matrix elements are
calculated using clover valence quarks. We examine the
following sources of systematic errors -- contribution of excited
states, estimates of renormalization constants, finite volume and
lattice discretization effects and dependence on quark mass. Three of
these sources, statistics, contribution of excited states, and
renormalization constants, affect the precision with which estimates
from an individual ensemble are extracted. The other three, finite
volume and lattice discretization effects and dependence on quark
mass, require fits and extrapolations based on all the points. We
examine the two classes of uncertainties separately.


\section{Statistics}
\label{sec:stat}

The MILC Collaboration~\cite{Bazavov:2012xda} has generated ensembles
of roughly 5500 trajectories of 2+1+1-flavor HISQ lattices at three
values of light quark masses corresponding to $M_\pi \approx 310$,
220, 130 MeV at $a=0.12$, $0.09$ and $0.06$ fm.  We analyze
configurations separated by 4--6 trajectories of the hybrid Monte
Carlo evolution and discard the initial 300--500 trajectories for
thermalization. The status of our analyses using these ensembles are
summarized in Table~\ref{tab:ensembles}.  To increase the statistics,
each configuration is analyzed using gaussian smeared sources in
multiple locations displaced both in time and space directions to
reduce correlations.

We performed the following statistical tests. The data for a given
ensemble are divided into bins (by source point and configurations)
and the Kolmogorov--Smirnov test is performed on quantities that have
reasonable estimates configuration by configurations (iso-vector
vector charge, value of 2-point function at a given time
separation). While, this pairwise test showed that the sub-samples are
consistent with being drawn from the same distribution, the mean
values of observables fluctuated by up to 3$\sigma$. This variation is
much larger than expected based on our bin size of over 1000
measurements. We do not find long tails in the distributions for any
of the samples that could explain the fluctuation, but do see a variation in
the sample distribution. One possible explanation is that the ensembles
have not covered enough phase space and errors are consequently 
underestimated. Our overall conclusion is that a few thousand,
and possibly $O(10,000)$ for the scalar charge, independent
configurations with $O(32)$ measurements on each are needed to obtain
estimates with $\le 2\%$ errors.

\begin{table*}
\centering
\begin{tabular}{|ccccccc|}
\hline
Label    & $L^3\times T$   & $M_\pi$ MeV  & $(M_\pi L)$ & $N_\text{cfgs}$ & $N_\text{Measurements}$ & $t_\text{sep}$ \\
\hline
a12m310  &$24^3\times 64$  & 305.3(4)     & $4.54$   &  1013   &  8104    & 8, 9, 10, 11, 12  \\
a12m220S &$24^3\times 64$  & 218.1(4)     & $3.22$   &  1000   &  24K (12K) & 10 (8, 12)         \\
a12m220  &$32^3\times 64$  & 216.9(2)     & $4.3$    &  958    &  7664    & 8, 10, 12         \\
a12m220L &$40^3\times 64$  & 217.0(2)     & $5.36$   &  1010   &  8080    & 10                \\
\hline
a09m310  &$32^3\times 96$  & 312.7(6)     & $4.5$    &  881    &  7058    & 10, 12, 14        \\
a09m220  &$48^3\times 96$  & 220.3(2)     & $4.71$   &  890    &  7120    & 10, 12, 14        \\
a09m130  &$64^3\times 96$  & 128.2(1)     & $3.66$   &  883    &  4824    & 10, 12, 14        \\
\hline
a06m310  &$48^3\times 144$ & 319.3(5)     & $4.51$   &  865    &  3460    & 16, 20        \\
a06m220  &$64^3\times 144$ & 229.2(4)     & $4.25$   &  430    &  1320    & 16, 20, 22, 24    \\
\hline
\end{tabular}
\caption{Description of the nine ensembles at $a=0.12$, $0.09$, $0.06$~fm used in this study.}
\label{tab:ensembles}
\end{table*}

\section{Excited-State Contamination}

Our current data show significant excited state contamination in both
2-point and 3-point functions.  The goal is to extract all observables
(charges, charge radii, form factors) by calculating matrix elements
between ground-state nucleons. We address this by using smeared
operators tuned to increase coupling to the ground state and suppress
radially excited states. Second, as discussed
in~\cite{Bhattacharya:2013ehc}, we partially remove the remaining
contamination by including one excited state in the analysis. Higher
states are not included because with current statistics we are not
able to resolve them, especially in the 3-point functions.

Denoting the first excited state mass by $M_1$ and coupling to our
operator by ${\cal A}_1$, the three-point function with source at
$t_i=0$, operator insertion at $t=t$ and sink at $t_f= t_{\rm sep}$
can be written as
\begin{align}
{\cal C}^{(3),T}_{\Gamma}(t_i,t,t_f;\vec{p}_i,\vec{p}_f) \approx \ &
       |{\cal A}_0|^2 \langle 0 | O_\Gamma | 0 \rangle  e^{-M_0 (t_f-t_i)} \ + \ 
       |{\cal A}_1|^2 \langle 1 | O_\Gamma | 1 \rangle  e^{-M_1 (t_f-t_i)} \nonumber\\
      +{} \ & {\cal A}_0{\cal A}_1^* \langle 0 | O_\Gamma | 1 \rangle  e^{-M_0 (t-t_i)} e^{-M_1 (t_f-t)} + {}\nonumber\\
      +{} \ & {\cal A}_0^*{\cal A}_1 \langle 1 | O_\Gamma | 0 \rangle  e^{-M_1 (t-t_i)} e^{-M_0 (t_f-t)} \,. 
\label{eq:three-pt}
\end{align}
The masses and amplitudes $M_0$, $M_1$, ${\cal A}_0$, and ${\cal A}_1$
are obtained from fits to the two-point functions. The desired matrix
element $\langle 0 | O_\Gamma | 0 \rangle$ is then obtained by
isolating $\langle 0 | O_\Gamma | 1\rangle $ and $\langle 1 | O_\Gamma
| 1 \rangle$. This requires doing calculations with multiple values of
$t$ and $t_{\rm sep}$. Using the sequential source method, we carry
out operator insertion at all values of $t$ between the source and
sink timeslices.  The values of $t_{\rm sep}$ investigated are listed
in Table~\ref{tab:ensembles}. A nonlinear least-square fitter is then
used to extract $\langle 0 | O_\Gamma | 0 \rangle$ by fitting the data
for all $t_{\rm sep}$ simultaneously using Eq.~(\ref{eq:three-pt}).

\begin{figure}
\includegraphics[width=.33\textwidth]{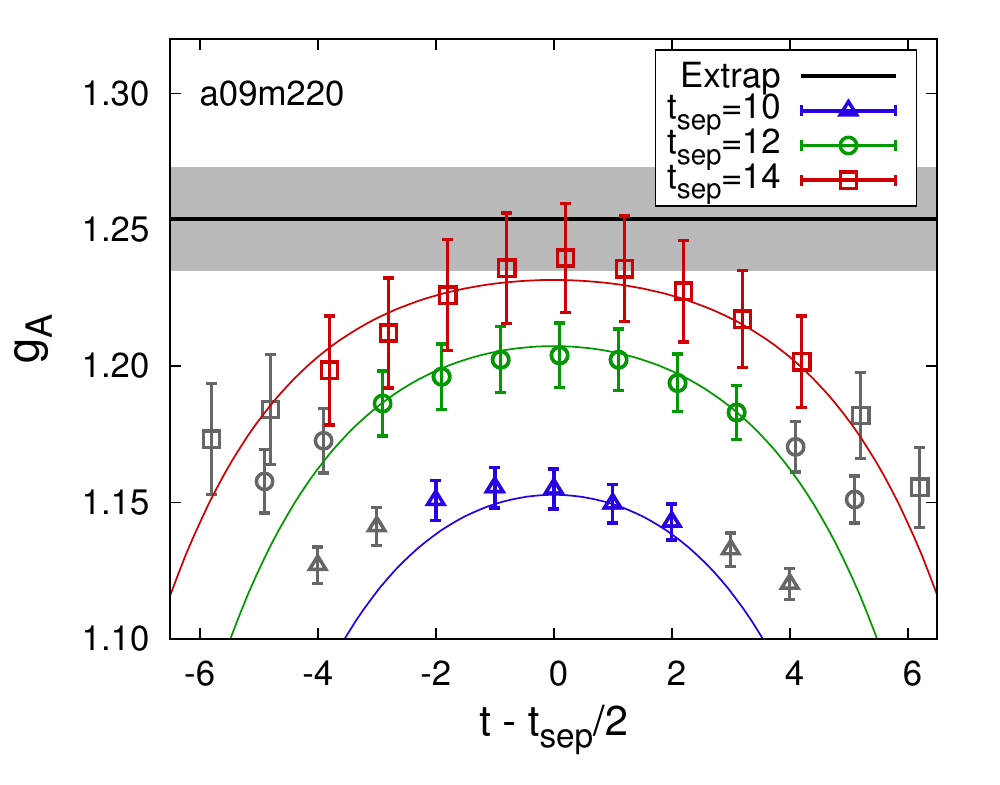} 
\includegraphics[width=.33\textwidth]{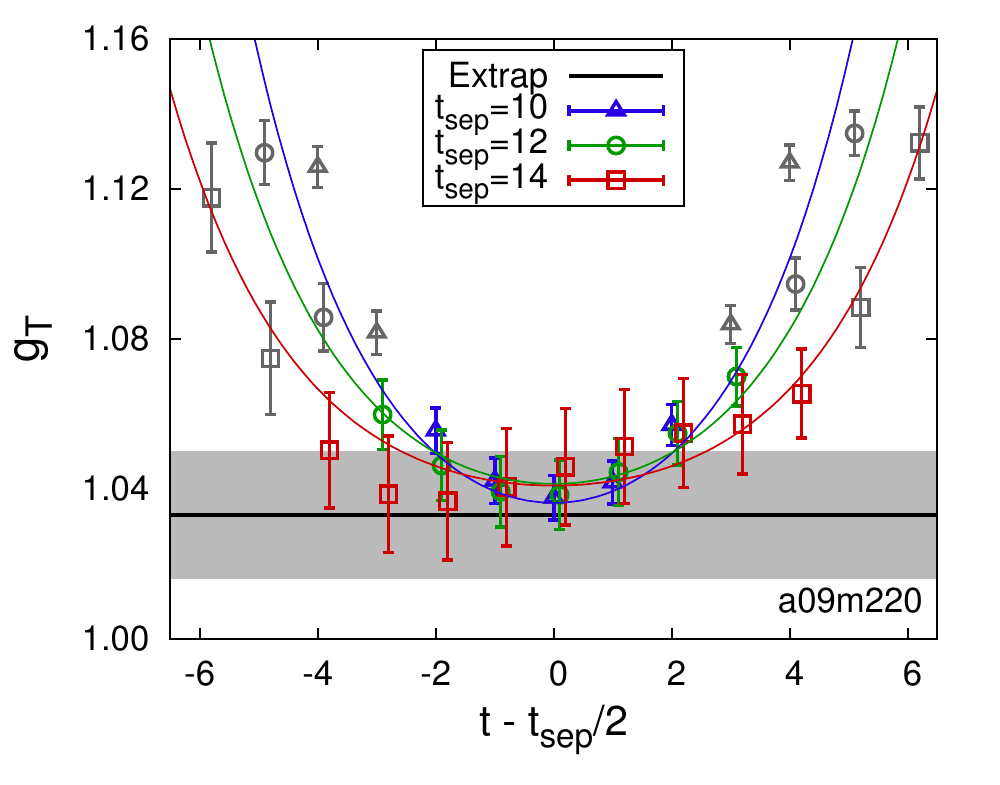} 
\includegraphics[width=.33\textwidth]{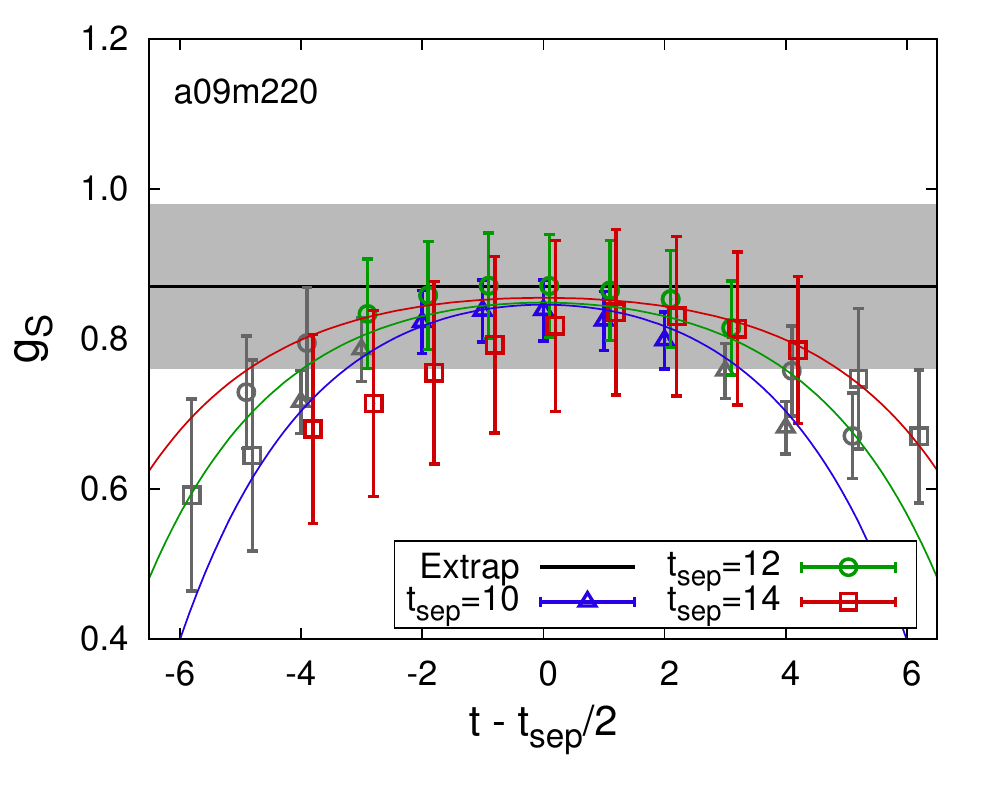} 
\vspace{-20pt}
\caption{Fit using Eq.(\protect\ref{eq:three-pt}) to extract
  unrenormalized $g_{A,S,T}$ from the a09m220 ensemble data. The black
  line and the grey error band are the result of the simultaneous fit
  to the $t_{\rm sep} = 10$, $12$ and $14$ data. The 3 colored
  lines are obtained from the simultaneous fit evaluated for each
  $t_{\rm sep}$.  To reduce excited-state contamination, the grey
  points on either side that are close to the source/sink timeslice, are
  not included in the fit.}
\label{Fig:a09m220Fits}
\end{figure}

The $a=0.12$ data for all three charges do not exhibit significant
trends with respect to $t_{\rm sep}$~\cite{Bhattacharya:2013ehc}.
Simultaneous fits to $t_{\rm sep}= 8,\ 10$ and $12$ data are
consistent with a fit to just the $t_{\rm sep}=10$ data. Consequently,
in~\cite{Bhattacharya:2013ehc} we had concluded that $t_{\rm sep} \ge
1.2$~fm is needed to control excited state contamination.  The
$a=0.09$ and $0.06$ data are much cleaner and show an increase in
$g_A$ with $t_{\rm sep}$ as illustrated in Fig.~\ref{Fig:a09m220Fits}.
On the other hand $g_T$ continues to show very little senstivity to
$t_{\rm sep}$. The errors in $g_S$ are too large to draw
conclusions. Overall, trends in $g_A$ data at weaker couplings imply
that $t_{\rm sep} \ge 1.5$fm is needed to establish control over
excited state contamination.

Our fits are biased by the smallest $t_{\rm sep}$ data because the
statistics are the same for all $t_{\rm sep}$, while errors increase
with $t_{\rm sep}$. For example, on the $a=0.12$fm ensembles, the
statistical errors increase by about $40 \%$ with each unit increase
in $t_{\rm sep}$.  This estimate scales with $a$, $i.e.$, on the
$a=0.06$fm ensembles, the same fractional increase takes place every 2
units. Similarly, the errors increase by about 20\% on lowering the
light ($u$ and $d$) quark masses by a factor of two, $i.e.$, going
from $M_\pi=310$ to 220 MeV ensembles. As an illustration, consider fits
to $a=0.09$fm ensembles with $t_{\rm sep} = 10,\ 12,\ 14$ shown in
Fig.~\ref{Fig:a09m220Fits}.  The $t_{\rm sep} = 10$ data make the
largest contribution to the extraction of $\langle 0 | O_\Gamma |
0\rangle $ and $\langle 0 | O_\Gamma | 1 \rangle$. The change between
$t_{\rm sep} = 10$ and $12$ contributes to fixing $\langle 1 |
O_\Gamma | 1\rangle $.  For $t_{\rm sep} = 14$ data to contribute at
the same level, its statistics should be 3--4 times that of $t_{\rm
  sep} = 10$ data.

\section{Finite Volume Effects}

\begin{figure}
\includegraphics[width=.33\textwidth]{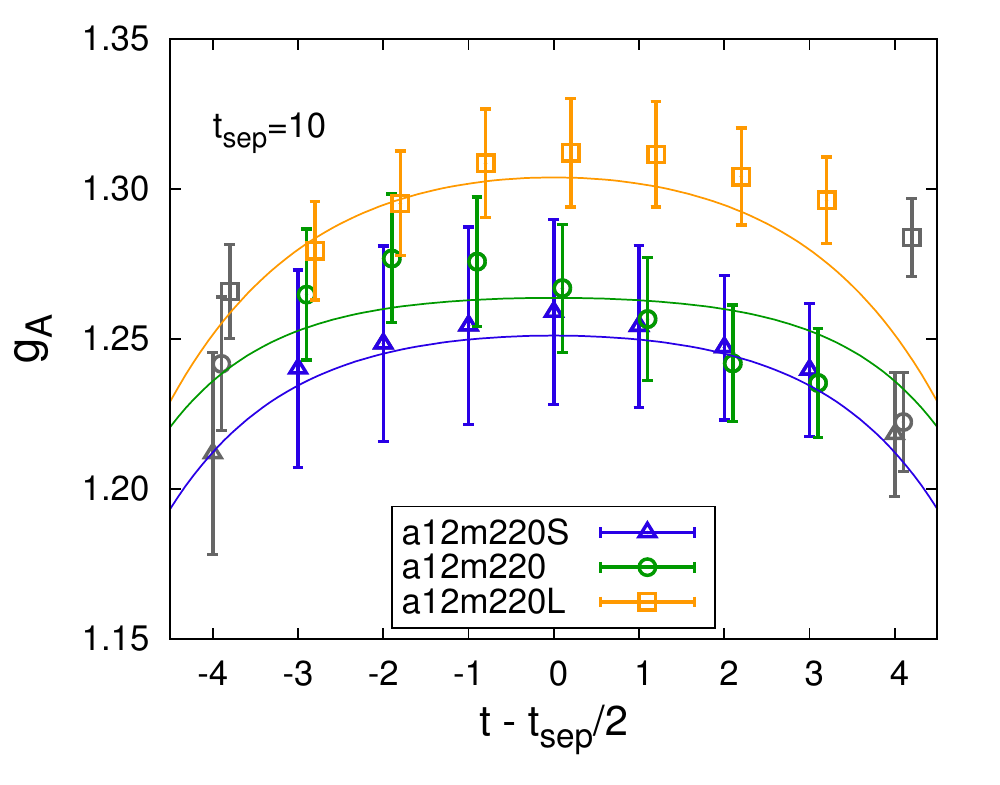} 
\includegraphics[width=.33\textwidth]{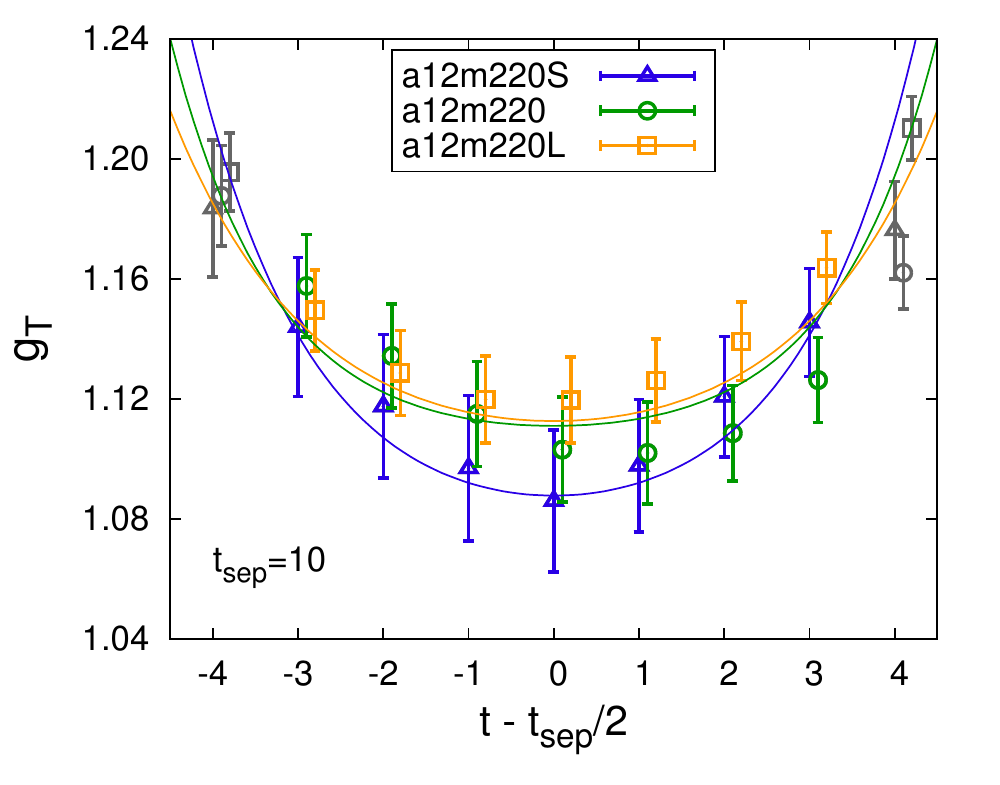} 
\includegraphics[width=.33\textwidth]{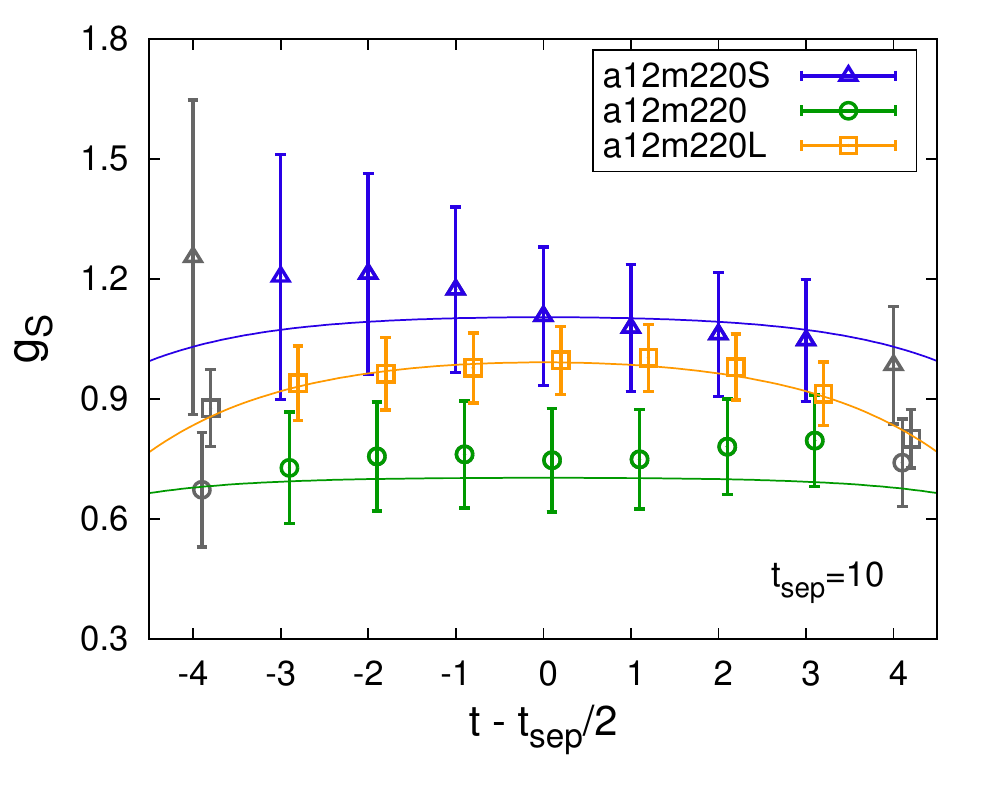} 
\vspace{-20pt}
\caption{Study of finite volume effects in unrenormalized $g_{A,S,T}$ using fits to
  Eq.(\protect\ref{eq:three-pt}) on the $t_{\rm sep} = 10$ data from
  the three a12m220 ensembles.  Rest is same as in
  Fig.~\protect\ref{Fig:a09m220Fits}.}
\label{Fig:a12m220Volume}
\end{figure}

The results of our finite volume study using the a12m220 ensembles
with volumes $24^3$, $32^3$ and $40^3$ (corresponding $M_\pi L =
3.22$, $4.3$ and $5.36$) are shown in Fig.~\ref{Fig:a12m220Volume}.
The $g_A$ data show significant increase with volume, while the $g_T$
data show saturation between the $M_\pi L = 4.3$ and $5.36$ data. The
$g_S$ data are again too noisy. Our conclusion is that lattices with 
$M_\pi L \ge 5$ are needed to control finite volume effects unless we can 
reliably model extrapolation in lattice volume.

\section{Non-perturbative Renormalization}

We use the RI-sMOM scheme to calculate the renormalization constants
of the bilinear quark operators~\cite{Martinelli:1994ty}. Details of
the method and our implementation were presented
in~\cite{Bhattacharya:2013ehc}.  The most important issue, especially
when using smeared lattices, was demonstrating the presence of a
window in momentum $q$, $ \Lambda_{QCD} \ll q \ll c/a$ with $c$ an
{\it a priori} unknown number of $O(1)$, where lattice artifacts are
expected to be small.  Sufficiently close to the continuum limit where
perturbation theory works, a test of whether such a window exists is
that $Z_S$ and $Z_T$ in the RI-sMOM (or any lattice) scheme should
show a $q^2$ dependence given by the anomalous dimensions of these
operators along with a weaker dependence from the running of
$\alpha_s$, while $Z_A$ should only show the latter. Converting
estimates obtained from within such a window in $q^2$ in the RI-sMOM
scheme to the $\overline{\rm MS}$ scheme used in pehnomenology and run
to some fixed scale, say $\mu=2$~GeV, should give estimates
independent of $q^2$.  Based on our analyses at the three lattice
spacings (see~\cite{Bhattacharya:2013ehc} for details) our conclusions
are: there is evidence of such behavior, $i.e.$, a window, in $Z_A$
and $Z_T$, but not in $Z_S$ even on $a=0.06$~fm ensembles. Lacking a
convincing demonstration of a window, we have defined a
procedure that will extrapolate to the right continuum
limit~\cite{Bhattacharya:2013ehc} and have been conservative in
estimating errors, however, we recommend a further study 
at various $a$, in particular for $Z_S$.

\section{Combined fits in lattice volume, spacing and quark mass}

\begin{figure}
\begin{center}
\includegraphics[width=.98\textwidth]{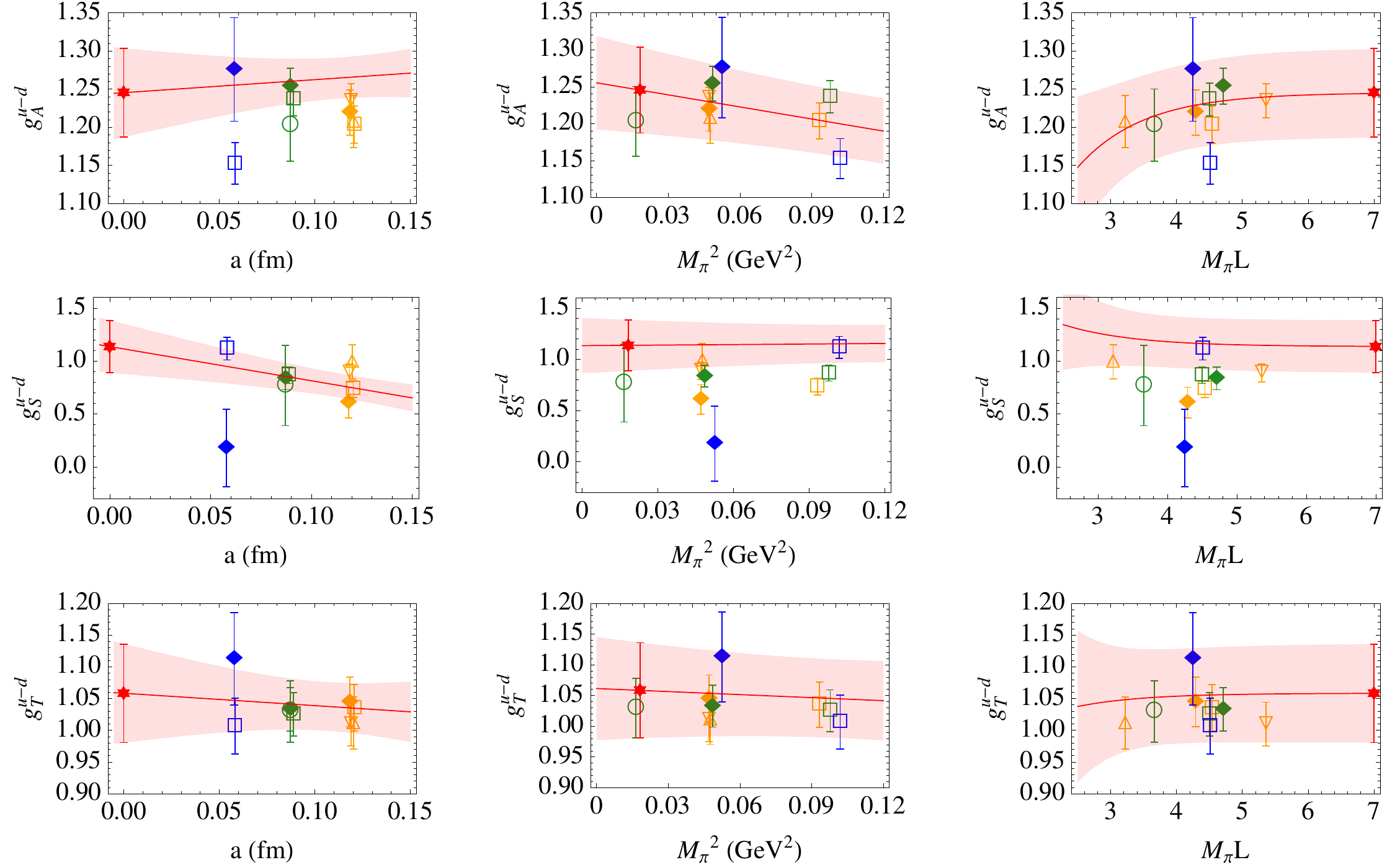}
\vskip -8pt
\caption{ Results of combined fits in $a$, $M_\pi$ and $M_\pi L$ to
  obtain the physical results for the renormalized charges using the
  ansatz given in Eq.~(\protect\ref{Eq:extrap}). Iso-vector charges are labeled as $g^{u-d}$.}
\label{fig:extrap}
\end{center}
\vskip -15pt
\end{figure}

Having discussed the first class of uncertainties that affect
individual data points, we now discuss extrapolations in lattice
spacing and volume, and the quark mass. It is very hard to generate
dynamical lattices with fixed quark masses (fixed $M_\pi$) and lattice
volumes (fixed $M_\pi L$) at multiple $a$ in order to take the
continuum limit along a line of constant physics. Similarly, it is not
easy to hold the lattice volume and $a$ constant and vary the quark
mass to study the chiral behavior.  Our best option is to do a
combined fit in $a$, $M_\pi$ and $M_\pi L$ to obtain physical
estimates. The second challenge is choosing the extrapolation ansatz
in each of these three variables --- we have to compromise between the
number of free parameters included and the number and quality of data
points.  In Fig.~\ref{fig:extrap} we show such a fit keeping only the
leading order terms in each of the three variables,
\begin{equation}
g(a, M_\pi, M_\pi L) = g^{{\rm physical}} + \alpha a + \beta M_\pi^2 + \gamma e^{-M_\pi L} \,.
\label{Eq:extrap}
\end{equation}
In Fig.~\ref{fig:extrap}, note that the errors in individual points
vary significantly.  As a result, with 9 data points, this ansatz with
4 free parameters is the most extensive we can explore.

Fig.~\ref{fig:extrap} summarizes the trends mentioned before. Removing
excited state contributions and doing finite volume and chiral
extrapolations all increase $g_A$ towards the experimental value. Data
for $g_T$ show almost no dependence on $a$, $M_\pi$ or $M_\pi L$ and
give $g_T=1.06 (0.06)$. We consider this estimate
reliable. Statistical errors are too large to draw conclusions
about $g_S$.

{\it Prognosis for the future}: Based on current analyses, we conclude that with
current ensembles and $O(10,000)$ measurements, we can extract $g_T$
with about 2\% errors on each point, and $\approx 5\%$ uncertainty in
the extrapolated value. To extract $g_A$ with similar precision will
require $O(2000)$ configurations at three values of $a \le 0.1$fm,
$M_\pi L \ge 5$ and $O(24)$ measurements on each configuration to get
statistically significant data for $1.2 \le t_{\rm sep} \le 1.6$fm.
Estimates of $g_S$ with similar precision will require an order of
magnitude more measurements.

\vskip -10pt
\section{BSM contributions to Neutron Electric Dipole Moment}

Lattice calculations of matrix elements of effective quark EDM and
chromo EDM operators within a neutron state to proble BSM theories
were initiated in~\cite{Bhattacharya:2012nEDM}.  The simpler is the
quark EDM which is an extension of $g_T$ but matrix elements of both
isovector and isoscalar tensor operators are need. One, therefore, has
to evaluate and control the signal in the disconnected
diagrams~\cite{Boram:Lattice2014}. We also analyze operator mixing and
renormalization in 1-loop perturbation theory. For brevity, operators
that vanish by the equations of motion are included by introducing the field
combinations:
\begin{eqnarray}
\psi_E & \equiv  & (i D^\mu \gamma_\mu - m) \psi~, \qquad \qquad \quad  \
 D_\mu = \partial_\mu - i g A_\mu^a T^a  - i e_\psi  A_\mu^{(\gamma)} \label{eq:defs1}
\\
\bar{\psi}_E  &\equiv & -  \bar{\psi}   \, (  i   \overleftarrow {D}^\mu \gamma_\mu  + m) ~, \qquad \qquad 
\overleftarrow {D}_\mu =   \overleftarrow{\partial}_\mu + i g A_\mu^a T^a +   i e_\psi  A_\mu^{(\gamma)}   ~.
\label{eq:defs2}
\end{eqnarray}
In terms of these fields, the operators we study are given in
Table.~\ref{Tab:nEDMops}.  The pattern of mixing of the dimension 5
operators under renormalization that needs to be calculated is given
in Table~\ref{Tab:mixing5}. Papers containing 1-loop results for the mixing and 
a first estimate of the quark EDM are being prepared.


\begin{table}
\begin{center}
\setlength{\tabcolsep}{12pt}
\begin{tabular}{|l|l|}
\hline
$O^{(3)}   =   i P =   \bar\psi i\gamma_5 \psi  $ & 
$O^{(5)}_1 =  C = \frac{ig}{2}\,  \bar\psi \left( \sigma^{\mu \nu} \gamma_5  + \gamma_5 \sigma^{\mu \nu} \right)  G_{\mu\nu}\psi  $
\\
 & 
$O^{(5)}_2 =   i  \partial^2  P $
\\
&
$O^{(5)}_3 =  E  = \frac{ie}{2}\,  \bar\psi \left( \sigma^{\mu \nu} \gamma_5  + \gamma_5 \sigma^{\mu \nu} \right)  F_{\mu\nu}\psi $
\\
$O^{(4)}_1 =  G \tilde{G} =  \frac{1}{2} \epsilon^{\mu \nu \alpha \beta}  G_{\mu \nu}^a  G_{\alpha \beta}^a $ &
$O^{(5)}_4  =    m \,  F \tilde{F} $
\\
$O^{(4)}_2 =  \partial  \cdot A   =   \partial_\mu  ( \bar\psi\gamma^\mu\gamma_5\psi) $ &
$O^{(5)}_5  =    m \,  G \tilde{G} $
\\
$O^{(4)}_3 =  i m P =  m  \bar\psi i\gamma_5 \psi $ &
$O^{(5)}_6  =   m  \, \partial \cdot A  $
\\
$O^{(4)}_4 =  F \tilde{F} =  \frac{1}{2} \epsilon^{\mu \nu \alpha \beta}  F_{\mu \nu}  F_{\alpha \beta} $ & 
$O^{(5)}_7   =    m^2 \, i  P $
\\
&
$O^{(5)}_8 =   i P_{EE}  =  i\bar\psi_E\gamma_5\psi_E $
\\
&
$O^{(5)}_9 =    \partial \cdot A_E  =  \partial_\mu[\bar\psi_E\gamma^\mu\gamma_5\psi+\bar\psi\gamma^\mu\gamma_5\psi_E] $
\\
&
$O^{(5)}_{10}  =  A_\partial  = 
  \bar\psi  \gamma_5  \slashed{\partial} \psi_E   \ -   \bar {\psi}_E   \overleftarrow{\slashed{\partial}}  \gamma_5 \psi  $
\\
&
$O^{(5)}_{11} =   A_{A^{(\gamma)}}  = i  e \left( \bar\psi\slashed{A}^{(\gamma)}  \gamma_5\psi_E - \bar\psi_E\slashed{A}^{(\gamma)}  \gamma_5\psi \right)$
\\
\hline
\end{tabular}
\end{center}
\vskip -10pt
\caption{Operators of dimension 3, 4 and 5 needed in the nEDM calculation.}
\label{Tab:nEDMops}
\end{table}


\begin{table}
\begin{center}
\setlength{\tabcolsep}{5pt}
\begin{tabular}{|c||c|c|c|c|c|c|c||c|c|c|c|}
\hline
& &   &  &  &   &   &  &      &          &  & 
\\
$ \ \  $ &  $C$& $\partial^2 P$   &  $E$ & $m F \tilde F$ & $m G \tilde G$ & $m \partial \cdot A$  &$m^2  P$  &  $P_{EE}$ &  $\partial \cdot A_E$  & $ A_\partial$ & $A_{A^{(\gamma)}}$  
\\[1\jot]
\hline
\hline
$C$& $Z_C$ &  $X$ & $X$ & $X$ & $X$  &$X$  &  $X$ &  $X$     & $X$         &  $X$ & $X$
\\
\hline
$\partial^2 P$&$0$ &  $Z_P$ & $0$ & $0$ & $0$  &$0$ &  $0$ &  $0$     & $0$         &  $0$ & $0$
 \\[1\jot]
\hline
$E$& $0$ &  $0$ & $Z_T$ & $0$ & $0$  &$0$  &  $0$ &  $0$     & $0$         &  $0$  & $0$
\\[1\jot]
\hline
$m F \tilde{F}$& $0$ &  $0$ & $0$            & $Z_m^{-1}  Z_{F \tilde F}$ & $0$  &$0$  &  $0$ &  $0$     & $0$         &  $0$ &  $0$ 
\\[1\jot]
\hline
$m\, G \tilde{G}$&$0$ &  $0$ & $0$ & $0$ & $Z_m^{-1} Z_{G \tilde G}$  &$X$  &  $0$ &  $0$     & $0$   &  $0$ & $0$
\\[1\jot]
\hline
$m \, \partial \cdot A$&$0$ &  $0$ & $0$ & $0$ & $0$  &$Z_m^{-1} Z_{\partial A}$  &  $0$ &  $0$     & $0$         &  $0$ & $0$
 \\[1\jot]
 \hline
 $ m^2 P$&$0$ &  $0$ & $0$ & $0$ & $0$  &$0$  &  $Z_m^{-1}$ &  $0$     & $0$         &  $0$ & $0$
  \\[1\jot]
\hline
\hline
$P_{EE}$& $0$ &  $0$ & $0$ & $0$ & $0$  &$0$  &  $0$ &  $X$     & $X$         &  $X$ & $0$
\\[1\jot]
\hline
$\partial \cdot A_E$& $0$ &  $0$ & $0$ & $0$ & $0$  &$0$  &  $0$ &  $0$     & $X$         &  $0$ & $0$
\\[1\jot]
\hline
$A_{\partial}$& $0$ &  $0$ & $0$ & $0$ & $0$  &$0$  &  $0$ &  $X$     & $X$         &  $X$&  $0$
\\[1\jot]
\hline
$A_{A^{(\gamma)}}$& $0$ &  $0$ & $0$ & $0$ & $0$  &$0$  &  $0$ &  $0$     & $0$         &  $0$ & $X$ 
 \\[1\jot]
\hline
\end{tabular}
\end{center}
\vskip -10pt
\caption{Mixing due to QCD of the dimension-5 operators. Non-zero entries need to be determined.}
\label{Tab:mixing5}
\end{table}

\vskip -30pt
\section*{Acknowledgments}
We thank the MILC Collaboration for sharing the 2+1+1 HISQ ensembles.
Simulations were performed using the Chroma software
suite~\cite{Edwards:2004sx} on LANL Institutional Computing, the USQCD
Collaboration facilities funded by the U.S. DoE; XSEDE supported by NSF grant
number OCI-1053575; and a DOE grant at NERSC. RG, TB and BY are supported
by DOE grant DE-KA-1401020 and the LDRD program at LANL. HL was
supported in part by DOE grant No. DE-FG02-97ER4014. The nEDM
calculations are being done  with V. Cirigliano and
E. Mereghetti.


\vskip -20pt

\begin{thebibliography}{99}

\bibitem{Bhattacharya:2011qm}
T. Bhattacharya, $et al.$, %
Phys.Rev. {\bf D85} (2012) 054512. 

\bibitem{Bhattacharya:2012nEDM}
T. Bhattacharya, $et al.$, %
PoS (LATTICE 2012), 179 (2012) and PoS (LATTICE 2013) 299 (2103).

\bibitem{Bazavov:2012xda}
A. Bazavov, et al., MILC Collaboration, Phys. Rev. {\bf D87}  (2013) 054505. 

\bibitem{Bhattacharya:2013ehc}
T. Bhattacharya, $et al.$, %
Phys.Rev. {\bf D89} (2014) 094502. 

\bibitem{Martinelli:1994ty}
G. Martinelli, et al., %
Nucl.Phys. {\bf B445} (1995) 81; 
C. Sturm, et al., 
Phys.Rev. {\bf D80} (2009) 014501. 

\bibitem{Boram:Lattice2014}
B. Yoon, et al., %
These proceedings: PoS (LATTICE 2014), 141 (2014).

\bibitem{Edwards:2004sx}
R. Edwards, B. Joo, Chroma Software System for LQCD, Nucl. Phys. Proc. Suppl. {\bf 140} (2005) 832

\end{thebibliography}
\end{document}